\documentclass[showpacs,aps,prl,twocolumn,superscriptaddress]{revtex4}
\usepackage{graphicx} 
\usepackage{dcolumn}
\usepackage{bm}
\usepackage{amssymb,amsmath}
\usepackage{epstopdf}
\usepackage[T1]{fontenc}
\usepackage[latin9]{inputenc}
\usepackage{color}
\usepackage{float}
\usepackage{graphicx}
\usepackage{esint}
\usepackage{bm}
\usepackage{graphics}
\makeatletter
\@ifundefined{textcolor}{}
{%
 \definecolor{BLACK}{gray}{0}
 \definecolor{WHITE}{gray}{1}
 \definecolor{RED}{rgb}{1,0,0}
 \definecolor{GREEN}{rgb}{0,1,0}
 \definecolor{BLUE}{rgb}{0,0,1}
 \definecolor{CYAN}{cmyk}{1,0,0,0}
 \definecolor{MAGENTA}{cmyk}{0,1,0,0}
 \definecolor{YELLOW}{cmyk}{0,0,1,0}
}

\begin{document}
\title{Effects of Electron-Electron Interactions on Electronic Raman Scattering of \\Graphite in High Magnetic Fields}
\normalsize
\author{Y.~Ma}
\thanks{Y.~Ma and Y.~Kim equally contributed to this work.}
\affiliation{Department of Physics and Astronomy, Rice University, Houston, Texas 77005, USA}
\author{Y.~Kim}
\thanks{Y.~Ma and Y.~Kim equally contributed to this work.}
\affiliation{National High Magnetic Field Laboratory, Tallahassee, Florida 32310, USA}
\author{N.~G.~Kalugin}
\affiliation{Department of Materials and Metallurgical Engineering, New Mexico Tech, Socorro, NM 87801, USA}
\author{A.~Lombardo}
\author{A.~C.~Ferrari}
\affiliation{Cambridge Graphene Centre, Cambridge University, 9 JJ Thomson Avenue, Cambridge, CB3 0FA, UK}
\author{J.~Kono}
\affiliation{Department of Physics and Astronomy, Rice University, Houston, Texas 77005, USA}
\affiliation{Department of Electrical and Computer Engineering, Rice University, Houston, Texas 77005, USA}
\author{A.~Imambekov}
\thanks{Deceased.}
\affiliation{Department of Physics and Astronomy, Rice University, Houston, Texas 77005, USA}
\author{D.~Smirnov}
\affiliation{National High Magnetic Field Laboratory, Tallahassee, Florida 32310, USA}

\date{\today}

\begin{abstract}
We report the observation of strongly temperature-dependent, asymmetric spectral lines in electronic Raman scattering of graphite in a high magnetic field up to 45~T applied along the $c$-axis. The magnetic field quantizes the in-plane motion, while the out-of-plane motion remains free, effectively reducing the system dimension from three to one. Optically created electron-hole pairs interact with, or shake up, the one-dimensional Fermi sea in the lowest Landau subbands. Based on the Tomonaga-Luttinger liquid theory, we show that interaction effects modify the van Hove singularity to the form $(\omega-\Delta)^{2\alpha-1/2}$ at zero temperature. We predict a thermal broadening factor that increases linearly with the temperature. Our model reproduces the observed temperature-dependent line-shape, determining $\alpha$ to be $\sim$0.05 at 40~T.
\end{abstract}

\pacs{78.30.-j, 71.70.Di, 73.61.Cw, 76.40.+b, 78.20.Bh}

\maketitle

Electron-electron interactions are progressively more important as the system dimension is lowered. One-dimensional (1d) systems, in particular, provide model environments in which to explore interaction effects~\cite{Giamarchi04Book}.  Interacting 1d electrons are expected to form an exotic electronic state of matter, the Tomonaga-Luttinger liquid (TLL)~\cite{Tomonaga50PTP,Luttinger63JMP,Haldane81PRL,Mahan00Book}.  A strong magnetic field, $B$, can suppress the electrons' kinetic energy, thus enhancing the effect of interactions, as exemplified by the fractional quantum Hall effect~\cite{TsuietAl82PRL,DuetAl09Nature,BolotinetAl09Nature}. In a 3d material, an applied magnetic field creates an effective 1d system along the field, ideal for a systematic study of interaction effects in a highly controllable fashion~\cite{BiaginietAl01EL}. Particularly promising are 3d metals with small electron and/or hole pockets near the Fermi energy ($E_{\rm F}$), such as bismuth~\cite{MaseSakai71JPSJ,NakajimaYoshioka76JPSJ,DuetAl05PRL,BehniaetAl07Science,LietAl08Science} and graphite~\cite{IyeetAl82PRB,YoshiokaFukuyama81JPSJ,KhveshchenkoetAl01PRL,DuetAl05PRL,KopelevichetAl09PRL}, where the magnetic quantum limit can be readily reached with $B\sim$10~T.

Here we use Raman spectroscopy to study electronic states and correlations in natural graphite in a strong $B$ up to 45~T, applied along the $c$-axis. This quantizes the electronic motion in the $ab$-plane, while the motion along the $c$-axis remains free, thus reducing the effective dimension from three to one. Instead of the main Raman features related to phonons~\cite{FerrarietAl06PRL,FerrariBasko13NN}, here  we focus on a series of electronic Raman features assigned to electronic inter-Landau-level (LL) transitions~\cite{KimetAl12PRB}, whose $B$-dependence can be explained through the Slonczewski-Weiss-McClure (SWM) model~\cite{McClure57PR,SlonczewskiWeiss58PR,McClure60PR}.  Each feature exhibits strongly temperature ($T$)-dependent shape. Our calculations show that scattering by thermally excited acoustic phonons~\cite{WoodsMahan00PRB,SuzuuraAndo02PRB,Jiang04CPL,MounetMarzari05PRB} is too weak to explain the observations. Electron-electron interactions, on the other hand, are shown to be the cause for the observed $T$ dependence, through the `shake-up' process known in the problem of X-ray (or Fermi-edge) singularities~\cite{Mahan00Book}. Namely, optically created electron-hole pairs interact with, or shake up, the 1d Fermi sea in the lowest Landau subbands, resulting in line-shape deviations from single-particle densities of states (i.e., 1d van Hove singularities).  Based on the TLL theory~\cite{Giamarchi04Book,Tomonaga50PTP,Luttinger63JMP,Haldane81PRL,Mahan00Book}, we show that electron-electron interactions modify the van Hove singularity to the form $(\omega-\Delta)^{2\alpha-1/2}$ at 0~K, where $\omega$ is the photon frequency, $\Delta$ the band-edge frequency, and $\alpha$ a dimensionless measure of the influence of electron-electron interactions. At finite $T$, we predict a thermal broadening factor, $\Gamma(T) \propto T$.  Our model reproduces the observed $T$-dependent line-shape, determining $\alpha$ to be 0.016, 0.026, and 0.048, at 20, 30 and 40~T, respectively.

Raman measurements are performed on natural graphite (NGS Naturgraphit GmbH) in a back-scattering Faraday geometry in $B$ up to 45~T, as described in Ref.~\onlinecite{KimetAl12PRB}.  The excitation beam from a 532-nm laser is focused to a spot size of $<$20~$\mu$m with a power of $\sim$13~mW.  Most of the data are collected with a spectral resolution of $\sim$3.4~cm$^{-1}$.  For the high-$B$, low-$T$ ($\leq$10~K) measurements of the sharpest peaks, a spectral resolution of $\sim$1.9~cm$^{-1}$ is used. The temperature drift over an integration time of up to 7 minutes, monitored by a temperature sensor installed below the sample, is $<$1~K at $T$ = 7~K and $<$2~K at $T\geq$~180~K.

\begin{figure}
\centerline{\includegraphics[scale=0.47]{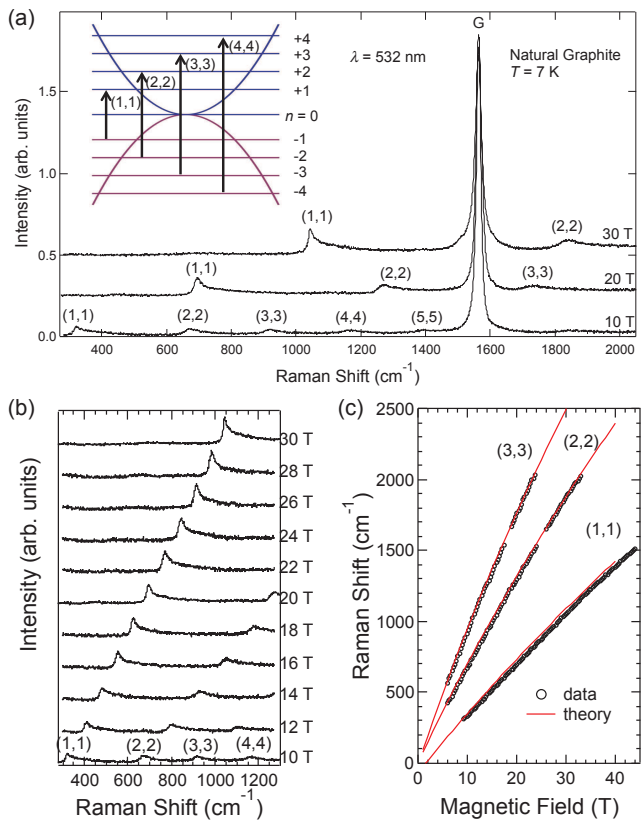}}
\caption{(Color online) (a)~Raman spectra at 10, 20, and 30~T.  The feature at $\sim$1580~cm$^{-1}$ is the G-peak due to the $E_{2g}$ phonons. Inset:~schematic energy level diagram showing the transitions responsible for the electronic peaks. (b)~Data taken at various $B$ at $T$ = 7~K, showing peaks due to (1,1) through (4,4) interband transitions. (c)~Peak positions of the observed (1,1), (2,2), and (3,3) transitions as a function of $B$, together with calculations based on the SWM model.}
\label{B-dependence}
\end{figure}

\begin{figure}
\centerline{\includegraphics[scale=0.52]{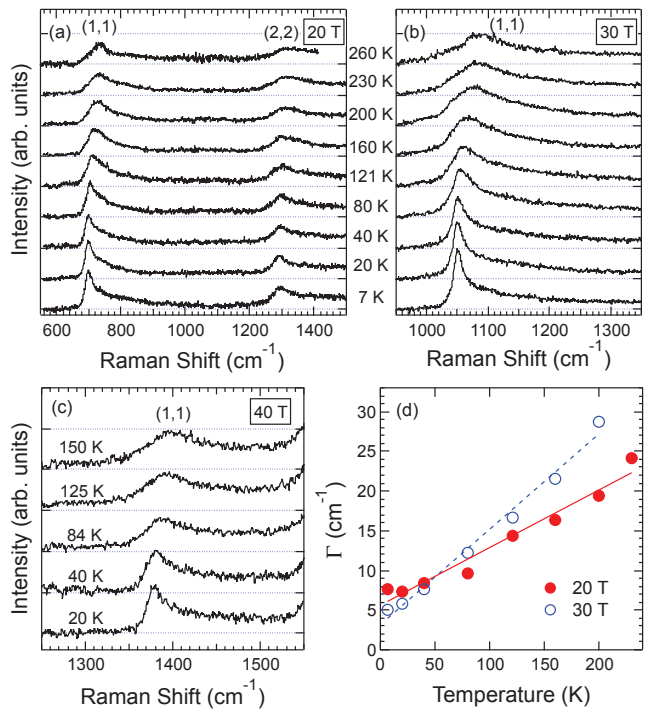}}
\caption{(Color online) T-dependent electronic Raman scattering of graphite at (a)20 (b) 30, and (c)40T. (d) T dependence of the broadening factor, $\Gamma$, of the (1,1) line at 20~T (solid circles) and 30~T (open circles). The lines are fits to the data}
\label{T-dependence}
\end{figure}

Figure \ref{B-dependence}(a) shows Raman spectra taken at 10, 20, and 30~T at 7~K.  The main feature is the G peak at $\sim$1580~cm$^{-1}$, due to $E_{2g}$ phonons~\cite{FerrariBasko13NN}. In the presence of $B$, {\em electronic} Raman features appear, coming from inter-LL transitions, labeled (1,1), (2,2), ... etc., which we focus on in this work. Figure \ref{B-dependence}(b) shows a series of spectra taken at various $B$ at 7~K, exhibiting electronic peaks that move with $B$.  These peaks can be attributed to the ``symmetric'' inter-LL excitations in the vicinity of the K point~\cite{KimetAl12PRB,KossackietAl11PRB}. The strongest, lowest-frequency transition, is (1,1), which is from the $n$ = $-$1 level in the valence band to the $n$ = 1 level in the conduction band. Similarly, we observe the (2,2), (3,3), and (4,4) transitions [see also the zero-field in-plane dispersions and energy levels near the K-point in the inset to Fig.~\ref{B-dependence}(a)].  The symmetric inter-LL excitations are non-resonant Raman processes and have been theoretically investigated for single-layer graphene (SLG)~\cite{KashubaFalko09PRB} and bilayer graphene (BLG)~\cite{MuchaetAl10PRB}.  The peak positions of the three lowest-energy transitions are plotted against $B$ in Fig.~\ref{B-dependence}(c); they agree well with our calculations (solid and dashed lines) based on the SWM model~\cite{KimetAl12PRB}.

These inter-LL transitions show strong $T$ dependence, as indicated in Fig.~\ref{T-dependence}, where Raman spectra at various $T$ are plotted for (a)~20, (b)~30, and (c)~40~T.  At the lowest $T$, the peaks exhibit sharp and asymmetric line-shapes, reminiscent of a 1d van Hove singularity, as expected from the effective dimension reduction from three to one in a $B$.  As $T$ increases, there is significant peak broadening and blue shift. The blue shift is expected from the thermal expansion of the carbon-carbon bonds, which changes the tight-binding parameters~\cite{MounetMarzari05PRB}. On the other hand, the thermal broadening cannot be explained within the tight-binding model. To quantify the thermal broadening, we first fit the spectra within a single-particle model using the joint density of states for interband transitions, obtained from the SWM model, with $T$-dependent Lorentzian broadening. Figure \ref{T-dependence}(d) shows the extracted Lorentzian full width at half maximum (FWHM) $\Gamma$ as a function of $T$ for 20 and 30~T. Apart from a small finite line-width at $T$ = 0, $\Gamma_{0} \approx$ 5~cm$^{-1}$, possibly due to disorder, $\Gamma$ linearly depends on $T$.

\begin{figure}
\centerline{\includegraphics [scale=0.54]{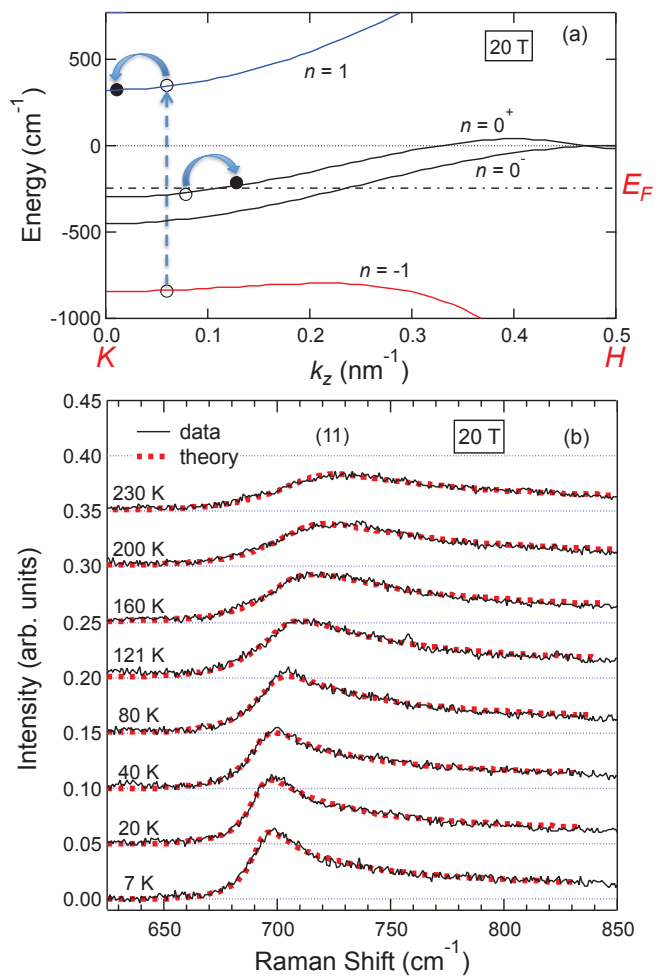}}
\caption{(Color online) (a)~Electron-electron interaction process influencing the electronic Raman line-shape of graphite in high B. The (1,1) electron-hole pairs interact with, or shake up, the 1d Fermi sea in the lowest-energy Landau subbands, creating particle-hole pairs across the Fermi energy. (b) T dependence of the line-shape for the (1,1) transition at 20~T, together with fits (dashed lines) based on the model in (a).}
\label{e-e model}
\end{figure}

Within the single-particle picture, $T$ only appears in the Fermi-Dirac distribution function, but this is a negligible effect since both the initial and final states of the Raman process are far away from $E_{\rm F}$, which resides in the $n$ = 0 bands. For example, for the (1,1) transition at 30~T, the electron and hole bands are $\sim$65~meV (or $\sim$750~K) away from $E_{\rm F}$. Thus, we need to take into account the interactions of the photo-excited electron-hole (e-h) pairs with some low-energy modes that would significantly change when $T$ changes from 4 to 300~K. Specifically, since the linear-$T$ broadening in Fig.~\ref{T-dependence}(d) implies a Bose-Einstein distribution at an energy scale much smaller than $k_{\rm B}T$, we only consider bosonic excitations whose characteristic energies are $\ll$~100~K.  Hence, we consider two types of low-energy modes: i)~particle-hole (p-h) excitations across $E_{\rm F}$ in the $n$ = 0 bands [Fig.~\ref{e-e model}(a)] and ii)~acoustic phonons (the lowest optical phonons being the shear modes, seen at $\sim$5-6~meV in graphite~\cite{tan}). We find that interactions with i) explain the observed $T$-linear broadening while interaction with ii) is too weak.

The magneto-electronic Raman scattering matrix was previously calculated for SLG~\cite{KashubaFalko09PRB} and BLG~\cite{MuchaetAl10PRB} and can be readily generalized to graphite in the presence of $B$:
\begin{equation}
\hat{R}=\Lambda\sum_{\vec{k}}\Psi_{n}^{\dagger}(k_{y},k_{z})\Psi_{-n}(k_{y},k_{z})
\label{eq:Raman-effective}
\end{equation}
where $\Lambda$ is the scattering amplitude, $k_y$ ($k_z$) are electron momenta in the $ab$-plane (along the $c$-axis), $\Psi_{n}^{\dagger}$ creates an electron in the $n$ = 1, 2, 3, 4, ... bands, and $\Psi_{-n}$ creates a hole in the $n$ =$-$1,$-$2, $-$3, $-$4, ... bands.  Both electrons and holes are massive at the bottom of the bands at the K point, i.e., $m_n \ne 0$ for all $n$'s, similar to BLG, but there is electron-hole asymmetry, i.e., $m_{1}\neq m_{-1}$.

Figure \ref{e-e model}(a) depicts the basic ingredients involved in the electron-electron interaction process we consider here, together with dispersions calculated via the SWM model for the $n$ = 0$^{\pm}$ and $\pm$1 bands at 20~T.  The two lowest-energy bands ($n$ = 0$^{\pm}$) cross $E_{\rm F}$, and the carriers near $E_{\rm F}$ have approximately linear dispersions. In the (1,1) process, an e-h pair is created in the $n = \pm1$ bands, which interact with, and are thereby dressed with, multiple p-h excitations in the $n$ = 0$^{\pm}$ bands near $E_{\rm F}$.  As $T$ is raised, the thermal smearing of the Fermi edge leads to stronger interaction between the massive e-h pair and the massless p-h pairs, and the peak broadens. This type of shake-up process was theoretically studied in carbon nanotubes at 0~K~\cite{Balents00PRB,MishchenkoStarykh11PRL}: a 1d van Hove singularity,  $(\omega-\Delta)^{-1/2}$, is predicted to become $(\omega-\Delta)^{-1/2+2\alpha}$ with $\alpha\sim0.1$ once the shake-up process is taken into account.

We describe the $n$ = 0$^{-}$ electrons as a Tomonaga-Luttinger liquid with the Hamiltonian~\cite{Giamarchi04Book,Tomonaga50PTP,Luttinger63JMP,Haldane81PRL}:
\begin{equation}
H_{0}^{\rm c}=v_{\rm F}\int dz\left [\psi_{\rm R}^{\dagger}i\partial_{z}\psi_{\rm R}-\psi_{\rm L}^{\dagger}i\partial_{z}\psi_{\rm L} \right ],
\label{eq:low energy mode of electron}
\end{equation}
where $v_{\rm F}$ is the Fermi velocity and $\psi_{\rm R(L)}^{\dagger}$ creates a particle near the right (left) Fermi point. The $n$ = 0$^+$ band can be described by a similar Hamiltonian, but with a different $v_{\rm F}$. By approximating the energy dispersion near $E_{\rm F}$ as $E \propto k_{z}$, we rewrite Eq.~(\ref{eq:low energy mode of electron}) via bosonization:
\begin{equation}
H_{0}^{\rm c} = \frac{v_{\rm F}}{2\pi}\int dz \left [(\nabla\phi)^{2}+(\nabla\theta)^{2} \right ],
\label{eq:bosonization of electrons}
\end{equation}
where $\nabla\phi=-2\pi[\rho_{\rm R}+\rho_{\rm L}]$, $\nabla\theta=2\pi[\rho_{\rm R}-\rho_{\rm L}]$, and $\rho_{\rm R}$ ($\rho_{\rm L}$) is the density operator for right-moving (left-moving) electrons.

We assume that the photogenerated electrons ($n$ = 1) and holes ($n=-1$) interact with the $n$ = 0$^-$ conduction electrons separately. For the $n$ = 1 band, where electrons are massive, we can treat the electrons through:
\begin{equation}
H_{1}=\int dz\Psi_{1}^{\dagger} \left [-\frac{1}{2m}\partial_{z}^{2}+\Delta_{1} \right ]\Psi_{1},
\label{eq:impurity hamiltonian}
\end{equation}
where $\Delta_1$ is the band edge frequency and $\Psi_{1}^{\dagger}$ ($\Psi_{1}$) is the creation (annihilation) operator for the $n$ = 1 band.  We also assume that the interaction Hamiltonian only involves the total charge density, thus neglecting any backscattering and Umklapp scattering:
\begin{equation}
H_{\rm int}=\frac{V}{2}\int dz \left [\Psi_{1}^{\dagger}\Psi_{1}-\frac{1}{2\pi}\nabla\phi \right ]^{2}.
\label{eq:interaction hamiltonian}
\end{equation}
We write the effective Hamiltonian for the system as the sum of Eqs.~(\ref{eq:bosonization of electrons})-(\ref{eq:interaction hamiltonian}): $H=H_{0}^{\rm c}+H_{1}+H_{\rm int}.$
Its diagonalization is a unitary transformation $U^{\dagger}HU$ and has been previously solved by many authors~\cite{Balents00PRB,MishchenkoStarykh11PRL,ImambekovGlazman09PRL,ImambekovGlazman09Science}:
\begin{equation}
U^{\dagger}=\exp \left [-i\frac{\gamma^{+}}{\pi}\int dy\theta(y)\Psi_{1}^{\dagger}(y)\Psi_{1}(y) \right ].
\label{eq:unitary transformation}
\end{equation}
Under this transformation, the original interacting system can be mapped to a non-interacting one, and the massive-electron operator acquires an additional string operator,
$\Psi_{1}(z)=\exp[-i\gamma^{+}\theta(z)/\pi]\tilde{\Psi}_{1}(z)$, where $\tilde{\Psi}^{\dagger}_1$ creates a free electron in the $n$ = 1 band.  The massive $n$ = 1 electron then gets dressed by the additional string operator, i.e., the $n$ = 0$^-$ conduction electrons adiabatically adjust to the massive electrons. Similarly, we can obtain a dressed expression for the massive hole.

The spectral function can be obtained by calculating the imaginary part of the retarded Green's function~\cite{Mahan00Book}:
\begin{equation}
G^{\rm R}(z,t)\equiv-i\theta(t)\langle[\Psi_{-1}^{\dagger}(z,t)\Psi_{1}(z,t),\Psi_{1}^{\dagger}(0,0)\Psi_{-1}(0,0)]\rangle.
\label{eq:green's function}
\end{equation}
At zero $T$, Eq.~(\ref{eq:green's function}) can be evaluated directly in real space. However, at finite $T$, one has to follow a different route. As the Green's function for the massive electron/hole and that for the conduction electrons are both straightforward to obtain, the total Green's function can be written as a convolution of three Green's functions:
\begin{eqnarray}
G^{\rm R}(z,t) & \approx & -i\theta(t)[-iG_{-1}^{<}(-z,-t)] \nonumber \\
&  & \times [iG_{1}^{>}(z,t)]F(z,t), \nonumber \label{eq:effective green's function} \\
G^{\rm R}(0,\omega) & = & -i\int\prod\frac{dp_{i}}{2\pi}\frac{d\omega_{i}}{2\pi}G^{0}(p_{2},\omega_{2})F(p_{1},\omega_{1})\nonumber \\
 &  & \times\delta(0-p_{1}-p_{2})\delta(\omega-\omega_{1}-\omega_{2}), \nonumber \\
G^{0}(p,\omega) & = & \int\frac{dp_{1}}{2\pi}\int\frac{d\omega_{1}}{2\pi}G_{1}^{>}(p_{1,}\omega_{1}) \nonumber \\
 & & \times G_{-1}^{<}(p_{1}-p,\omega_{1}-\omega),
\label{eq:green's function as convolution}
\end{eqnarray}
where
\begin{equation}
F(z,t)  =  \langle\exp[-i\gamma\theta(x,t)]\exp[i\gamma\theta(0,0)]\rangle.
\end{equation}
We can express the spectral function in a universal form:
\begin{equation}
A(\omega)  =  \Lambda T^{2\alpha-0.5}\tilde{F}\left (\frac{\omega/T}{4\pi},\alpha \right ),
\end{equation}
where
\begin{eqnarray}
\tilde{F}(z,t) & = & \sum_{n=0}^{\infty}\sum_{m=0}^{\infty}B[n+\alpha,1-\alpha]B[m+\alpha,1-\alpha]\nonumber \\
 &  & \times {\rm Re}\left [ \frac{(2i)^{2\alpha}}{\sqrt{z-\frac{i}{2}(m+n+\alpha)}} \right ].
 \label{eq:final result}
\end{eqnarray}

In Eq.~(\ref{eq:final result}) there are two dimensionless parameters: $\omega/T$ and $\alpha$. The first implies that the spectral width linearly depends on $T$ for a fixed $\alpha$. The second can be understood by studying the $T$ = 0 asymptotic behavior of Eq.~(\ref{eq:final result}), and comparing it with the previous zero-$T$ results~\cite{Balents00PRB,MishchenkoStarykh11PRL}. It then becomes clear that:
\begin{equation}
A(\omega)\propto\frac{\Theta(\omega-\Delta)}{(\omega-\Delta)^{1/2-2\alpha}},
\label{eq:final prediction for dummy}
\end{equation}
where $\Theta$ is the Heaviside function. For metallic nanotubes, $\alpha$ was estimated to be$\sim0.1$~\cite{Balents00PRB,MishchenkoStarykh11PRL}. To fit our experimental data with our model, we use the true band structure instead of a parabolic approximation, by fitting the tail up to $\sim0.2(\pi /c)$ from the K point. Fig.\ref{e-e model}(b) shows how well our model fits the data, giving $\alpha$(20\,T) $\approx$ 0.016, $\alpha$(30\,T) $ \approx$ 0.026, and $\alpha$(40\,T) $\approx$ 0.048.

We now consider the longitudinal acoustic (LA) phonons in graphite, which can also couple to the massive electrons and holes. The properties of acoustic phonons can be described by five elastic constants~\cite{BosaketAl07PRB}: $C_{11}$ = 1109~GPa, $C_{66}$ = 485~GPa, $C_{33}$ = 38.7~GPa, $C_{13}$ = 0~GPa, and $C_{44}$ = 5~GPa. Unlike the case of optical phonons~\cite{SuzuuraAndo02PRB,Ando06JPSJ2,FerrarietAl06PRL,piscanec}, coupling with acoustic phonons vanishes at the $\Gamma$ point since $H_{\rm ep} \sim \sqrt{q}$~\cite{WoodsMahan00PRB,BoninietAl07PRL,GiuraetAl12PRB}, where $q$ is the phonon wavenumber. We then evaluate the thermal broadening of Raman peaks by calculating the imaginary part of the self-energy:
\begin{eqnarray}
H_{\rm ep} & = & \sum_{\vec{k}^{\prime},\vec{k},\vec{q}}g_{\vec{q}}h(\vec{q})\Psi_{1}^{\dagger}(k_{y}+q_{y},k_{z}+q_{z})\Psi_{1}(k_{y},k_{z})(b_{-\vec{q}}^{\dagger}+b_{\vec{q}})\nonumber \\
g_{\vec{q}} & = & \frac{\eta\kappa q\sin\theta}{2}\sqrt{\frac{\hbar}{2NM\omega_{\vec{q}}}}\frac{\sqrt{2}}{2}\frac{\Delta_{B}^{2}}{\Gamma\gamma_{1}}\label{eq:electron-phonon interaction}\\
h(\vec{q}) & = & \left ( 4 - 2l_{B}^{2}q^{2}\sin^{2}\theta + \frac{l_{B}^{4}q^{4}\sin^{4}\theta}{8} \right ) e^{-(l_{B}^{2}q^{2}\sin^{2}\theta) / 4} \nonumber
\end{eqnarray}
where $l_B = (\hbar / eB)^{1/2}$ is the magnetic length, $\eta\sim2$, and $\kappa\sim1/3$~\cite{SuzuuraAndo02PRB}. To first order, we estimate the scattering rate through Fermi's golden rule:
\begin{equation}
W_{i}=\frac{2\pi}{\hbar}\sum_{f}|\langle f|H_{\rm ep}|i\rangle|^{2}\delta(E_{i}-E_{f}).
\label{eq:scattering rate}
\end{equation}
When the momentum transfer during the scattering process is small (i.e., $vq\ll kT$), the phase space for phonon modes are $q^{2}(\frac{1}{e^{vq/kT}}+\frac{1}{2}\pm\frac{1}{2})\sim qT\to0$, and when the momentum transfer is large, the overlap between initial and final states has a factor $\exp(-q_{\perp}^{2}l_{B}^{2})$. For $B$ = 30~T, $l_B \sim$ 5~nm, which is at least one order larger than the carbon-carbon bond length. Thus, the contribution to scattering drops exponentially as the phonon modes move away from the $\Gamma$ point (or, equivalently, with increasing energy). The calculated momentum-dependent scattering rate is then given by
\begin{eqnarray}
W(k_{z}) & = & \Lambda^{\prime}\int_{0}^{\pi}d\theta\frac{\tilde{q}^{2}\sin^{3}\theta}{\sqrt{\sin^{2}\theta+\frac{V_{4}^{2}}{V_{1}^{2}}\cos^{2}\theta}}\frac{h^{2}(q,\theta)}{\cos^{2}\theta}\nonumber \\
 &  & \times \left (n_{\omega_{\vec{q}}}+\frac{1}{2}\pm\frac{1}{2} \right ),
\end{eqnarray}
where
\begin{eqnarray}
\tilde{q} & = & \frac{2ml_{B}V_{1}}{\hbar}\frac{\frac{\hbar k_{z}}{mV_{1}}\cos\theta\mp\sqrt{\sin^{2}\theta+V_{4}^{2}/V_{1}^{2}\cos^{2}\theta}}{\cos^{2}\theta}, \nonumber\\
\hbar\Lambda^{\prime} & = & \frac{\eta^{2}\kappa^{2}}{16\pi}\frac{m}{M}\frac{V_{\rm unit}}{l_{B}^{3}}\frac{\Delta_{B}^{2}}{\gamma_{1}^{2}}\frac{\sqrt{2}v}{V_{1}}\Delta_{B}
 \approx  4.1\times10^{-6} \, \, {\rm cm}^{-1}. \nonumber
\label{eq:imaginary part of self energy}
\end{eqnarray}
This value leads to $W(k_{z}) \sim$ 10$^{-4}$~cm$^{-1}$ at 30~T and 200~K, too small to explain the observed broadening, which requires the scattering rate to be $\sim$10~cm$^{-1}$.  There are two reasons for the small $\hbar\Lambda^{\prime}$ value: one is $m/M \sim 10^{-3}$, and the other $V_{\rm unit} / l_{B}^{2}$. The latter, i.e., the magnetic length suppression, is a unique aspect of this work, made possible by a high $B$.

In summary, we studied electronic Raman scattering in graphite in a strong magnetic field up to 45~T, applied along the $c$-axis. We observe a series of spectral lines, ascribed to inter-Landau-subband transitions, and each line exhibits strongly $T$-dependent line-shape. We developed a microscopic model based on the Tomonaga-Luttinger theory, with which we show that electron-electron interactions can explain the observed results, through the `shake-up' process known in the problem of X-ray (or Fermi-edge) singularities. Specifically, we show that electron-electron interactions modify the van Hove singularity to the form $(\omega-\Delta)^{2\alpha-1/2}$ at $T$ = 0.  Our model accurately reproduces the observed $T$-dependent line-shape, determining $\alpha$ to be 0.016, 0.026, and 0.048, at 20, 30, and 40~T, respectively. 

This work was supported in part by NHMFL UCGP-5068 and DOE/BES DE-FG02-07ER46451.  J.K.~acknowledges support from NSF (Grant No.~DMR-1006663), DOE BES Program (Grant No.~DE-FG02-06ER46308), the Robert A.~Welch Foundation (Grant No.~C-1509).  A.C.F.~thanks the Royal Society, the European Research Council Grant NANOPOTS, EU Grants RODIN, MEM4WIN, and CareRAMM, EPSRC grants EP/K01711X/1, EP/K017144/1, EP/G042357/1, the Leverhulme Trust, and Nokia Research Centre, Cambridge for support.

\end{document}